\documentclass[12pt]{iopart}
\usepackage{color}
\usepackage{graphicx}

\begin{document}

\title[Facets of glass physics]{Facets of glass physics}

\author{Ludovic Berthier$^1$ \& Mark D. Ediger$^2$}

\address{$^1$Laboratoire Charles Coulomb, UMR 5221, CNRS and 
Universit\'e Montpellier, Montpellier, France}

\address{$^2$Department 
of Chemistry, University of Wisconsin-Madison, Madison, Wisconsin 53706, USA}

\begin{abstract}
Glasses constitute a widespread form of solid matter, and glass
production has been an important human technology for more
than 3000 years. Despite that long history, new ways to understand
the fundamental physics of glasses continue to emerge.
\end{abstract}

The simplest way to make glass is to cool a material from its liquid
state quickly enough that nucleation and growth of a crystalline
phase do not happen. When the transition to the crystal is avoided,
particle motion inside the supercooled liquid slows dramatically
with decreasing temperature. Glassblowers use that slowdown to
shape the cooling material as its viscosity increases.
When the temperature decreases further, the time scale for 
molecular rearrangements becomes so long that the system falls out of
equilibrium with respect to the supercooled liquid state. The 
material stops flowing and becomes a glass~\cite{mark_review}. 
The temperature at which
that occurs is called the glass transition temperature $T_g$.

Amorphous solids can be made of atoms, simple organic molecules, 
larger molecules such as polymers, or assemblies of colloidal
particles. Even macroscopic constituents such as sand piles and
shaving cream may form rigid and disordered particle assemblies
that are analogous to molecular and atomic glasses. Glass formation
in different types of materials exhibits remarkable universal features. 
For example, only minor changes in the local structure accompany the huge 
increase in viscosity as temperature is decreased~\cite{mark_review}.
For many materials, producing a glass is not difficult, and cooling at a 
few kelvin per minute is sufficient to avoid crystallization.
The examples of ancient glasses described in Figure 1
demonstrate that metastability with respect to the crystal can be
readily achieved. However, for materials in which crystal nucleation is 
efficient, such as simple metals or highly symmetric molecules, glass 
formation can be challenging. Extremely fast cooling is
required, and in some cases, cooling rates as high as $10^9$~K/s may
not suffice.

\begin{figure}
\begin{center}
\includegraphics[width=13cm]{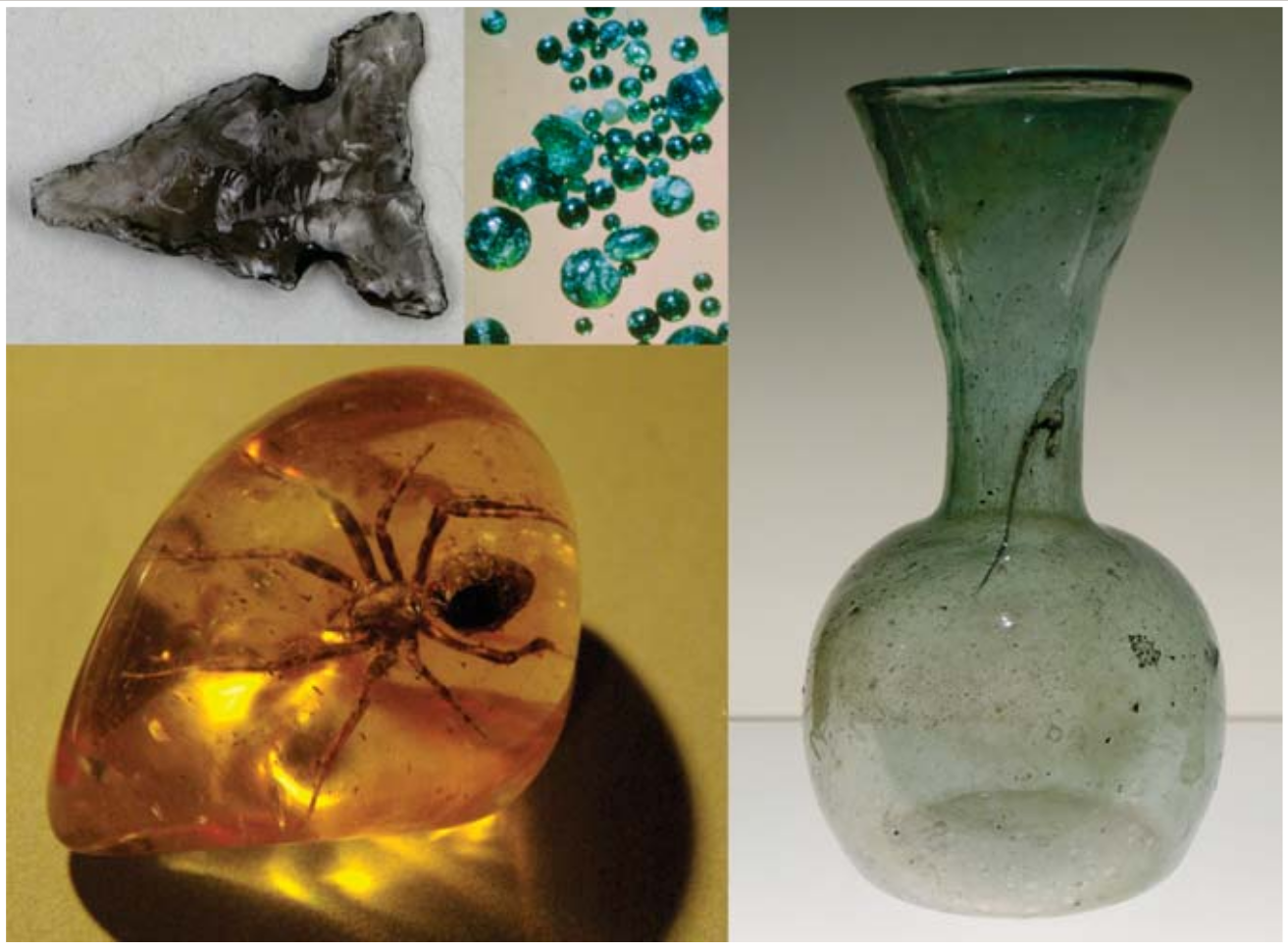}
\end{center}
\caption{{\bf The paradox of old glasses.}
Glasses are far-from-equilibrium materials produced by avoiding a 
thermodynamic transition to an ordered crystalline structure. Given enough
time, a glass will eventually reach its true equilibrium state: the crystal.
Remarkably, ancient glasses exist; the figure shows several examples.
The Egyptian jug (far right) is about 2500 years old. Researchers have
turned to 20-million-year-old amber glasses to learn how extended 
periods of aging influence glass properties~\cite{amber}. 
Amber glasses may encapsulate dinosaur-age traces of life (bottom left), 
as popularized by 
Hollywood blockbusters. Many early cultures fashioned arrowheads (top left)
and other tools from obsidian, a naturally occurring volcanic glass that
can remain glassy for 75 million years without crystallizing. Apollo 15
astronauts brought back small glass beads (center top) from the Moon.
Dated to be more than 3 billion years old, they are the oldest glasses
on Earth.
The long-term stability of some glasses might be surprising given
their nonequilibrium nature. That stability, and their flexibility with
regard to composition, makes glasses appealing as storage media for
nuclear waste. (See the article by Ian Pegg, Physics Today, February 2015,
page 33.) Storage at temperatures much less than the glass transition
temperature enhances glass stability. The low temperature slows the
molecular rearrangements that might lead to the nucleation and growth
of crystals.
Environmental factors are also important for glass stability. 
Geological glasses often
crystallize as a result of exposure to water. An oxidizing environment can
chemically change a glass and cause crystallization. On Earth, the slow
but persistent movement of tectonic plates recycles geological glasses
into other materials. Given the temperature and environment 
considerations, the Moon turns out to be nearly the perfect home for glasses.}
\label{fig1}
\end{figure}

In terms of molecular organization, glassy materials appear to
be the continuation of the liquid state. Particle configurations
closely resemble the disordered structure of the supercooled liquid,
as shown in Figure 2. However, glasses do not easily deform, and
thus they constitute a solid form of matter. Solidity, which is often
presented in solid-state physics textbooks as a direct consequence
of the broken translational invariance of periodic crystals, can exist
in fully aperiodic structures. Some molecular rearrangements and
flow can still occur below $T_g$ but only extremely slowly, the more
so the lower the temperature.
As macroscopically homogeneous solids, glasses are the best
materials for many applications, from optical fibers to windows. In
contrast, large crystals tend to have grain boundaries that scatter
light. In addition, glass composition can be widely varied to optimize 
properties. Macroscopic homogeneity and compositional flexibility result 
from the locally disordered, liquid-like structure of the
glass.

\begin{figure}
\begin{center}
\includegraphics[width=13cm]{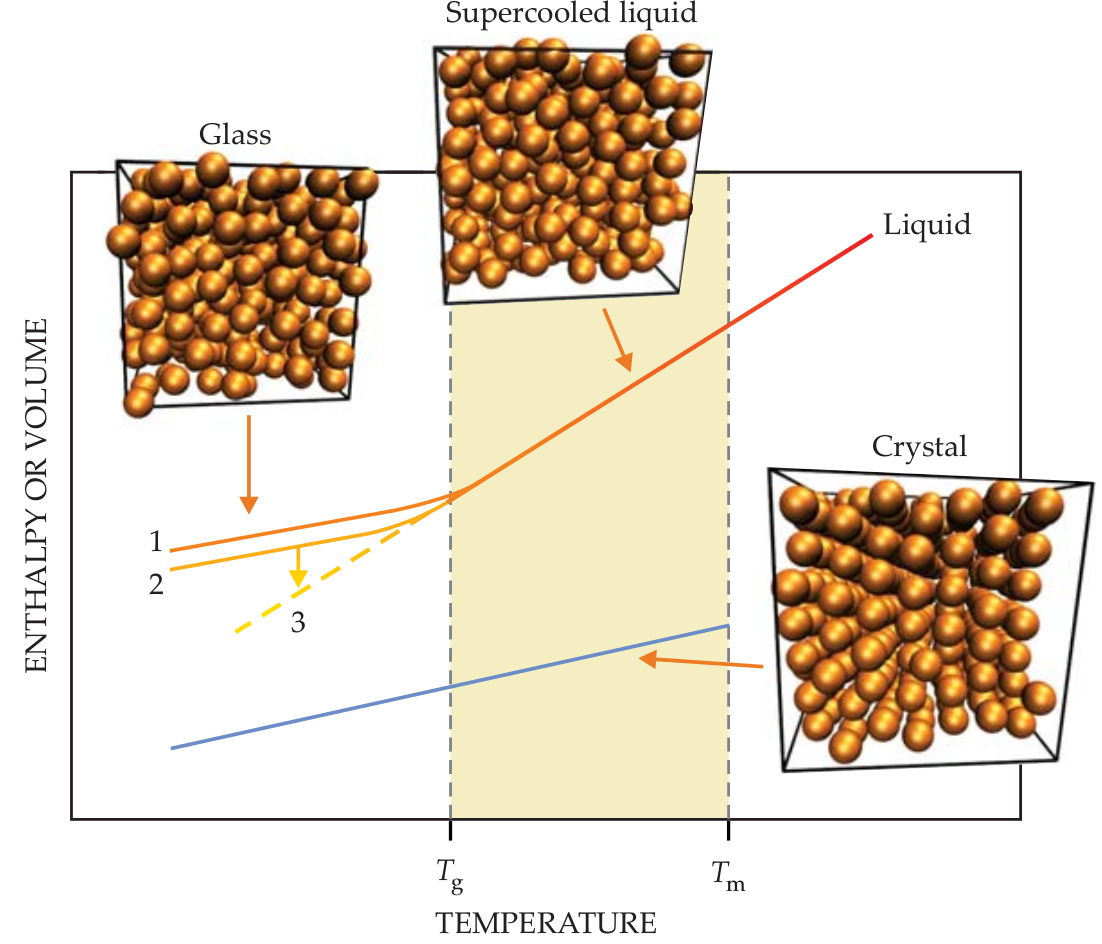}
\end{center}
\caption{{\bf Can you find the glass?} The enthalpy or molar volume of a
liquid as the temperature is lowered illustrates the production of three
different glasses. If crystallization is avoided, glass will be formed upon
cooling a liquid (glass 1). Slower cooling produces a denser glass (glass
2). Isothermal aging below $T_g$ produces an even denser glass (glass 3).
Typical configurations from molecular dynamics computer simulations
are shown for a glass, a supercooled liquid, and a crystal.}
\label{fig2}
\end{figure}

Glasses are also fascinating materials from a fundamental perspective 
because they represent nonequilibrium disordered states
of matter. Understanding them in the context of thermodynamic
phase transitions is an exciting challenge that has yet to be 
solved~\cite{rmp}.
The glass problem differs fundamentally from its crystalline sibling
whose configuration space is dominated by a unique free-energy
minimum that corresponds to a perfectly ordered structure. By contrast, 
a glass may take on a large number of equivalently disordered,
imperfect structures. When cooled, the exploration of that complex
free-energy landscape slows and the system must eventually
choose one of the many available glassy states. The final state of the
system is one among a vast library of possible disordered states,
each representing a local free-energy minimum.

\section*{How to build a better glass}

We often want to produce better glasses, where “better” might
mean stiffer, more resistant to impact, stable to higher temperatures, 
or some other desired property. The glass scientist has two
general approaches to optimize such properties.
The first approach is to vary the composition; glasses, in contrast to 
crystals, are quite flexible in that regard. The exploration of
different glass compositions is a major endeavor in the field of
metallic glasses (see the article by Jan Schroers, Physics Today, 
February 2013, page 32).

The second approach to optimize glass properties is to control
the preparation route for glass formation. Even with a fixed composition, 
the glassmaker can, in principle, produce an extremely
large number of distinct glasses by varying the formation process.
Figure 2 describes routes to three glassy
states. The sensitivity to preparation is a direct consequence of the
nonequilibrium nature of the glassy state. A good analogy is with
cooking: Different chefs come up with wildly different results even
when using the same set of ingredients.

Tempered glass is produced by cooling the surface of the glass
more rapidly than the interior. Because of the internal stresses that
get built in during formation, tempered glass explodes into many
small pieces when it is broken, whereas a slowly cooled glass with
identical composition would break into a handful of large, dangerous shards. 
Preparation thus may affect the mechanical properties
of the glass in a spectacular manner. The glass faces of smartphones
have been optimized in both composition and processing to robustly protect 
the display.

Suppose we want to make a higher-density glass in the expectation that 
it would also show improved stiffness and thermal stability. Figure 2
shows one method: If held for a long period of time
below $T_g$, the volume of a glass will decrease toward the supercooled 
liquid volume in a process known as physical aging. That
densification occurs as a result of the thermodynamic driving force
to reach the supercooled liquid state. However, the process is kinetically 
hindered by the extremely long times required for rearrangements of the 
local structure. The situation is actually worse because
for every bit of progress made toward the supercooled liquid, a
structure is formed that has even higher barriers to rearrangement.
Thus the process of densification through physical aging slows 
logarithmically over time~\cite{aging}.

Alternatively, researchers have recently discovered that high
density glasses can be assembled much more quickly in a process
known as physical vapor deposition~\cite{ultrastable}, 
as illustrated in Figure 3. 
Mobility at the free surface of a glass can be $10^9$ times higher than in
the interior and is the key to the process. Even below $T_g$, molecules
near the surface quickly equilibrate toward the supercooled liquid
state, a process that occurs much more slowly in the bulk. Much
like a block-stacking video game, further deposition locks the packing 
into place.
Surface mobility coupled with slow deposition leads to well-packed 
glasses that exhibit high kinetic stability, even at temperatures 
above $T_g$. Ultrastable glasses made via physical vapor 
deposition can be nearly 1.5~\% denser than conventional glasses.
Achieving such density through physical aging would require an
estimated $10^6$ years.

\begin{figure}
\begin{center}
\includegraphics[width=13cm]{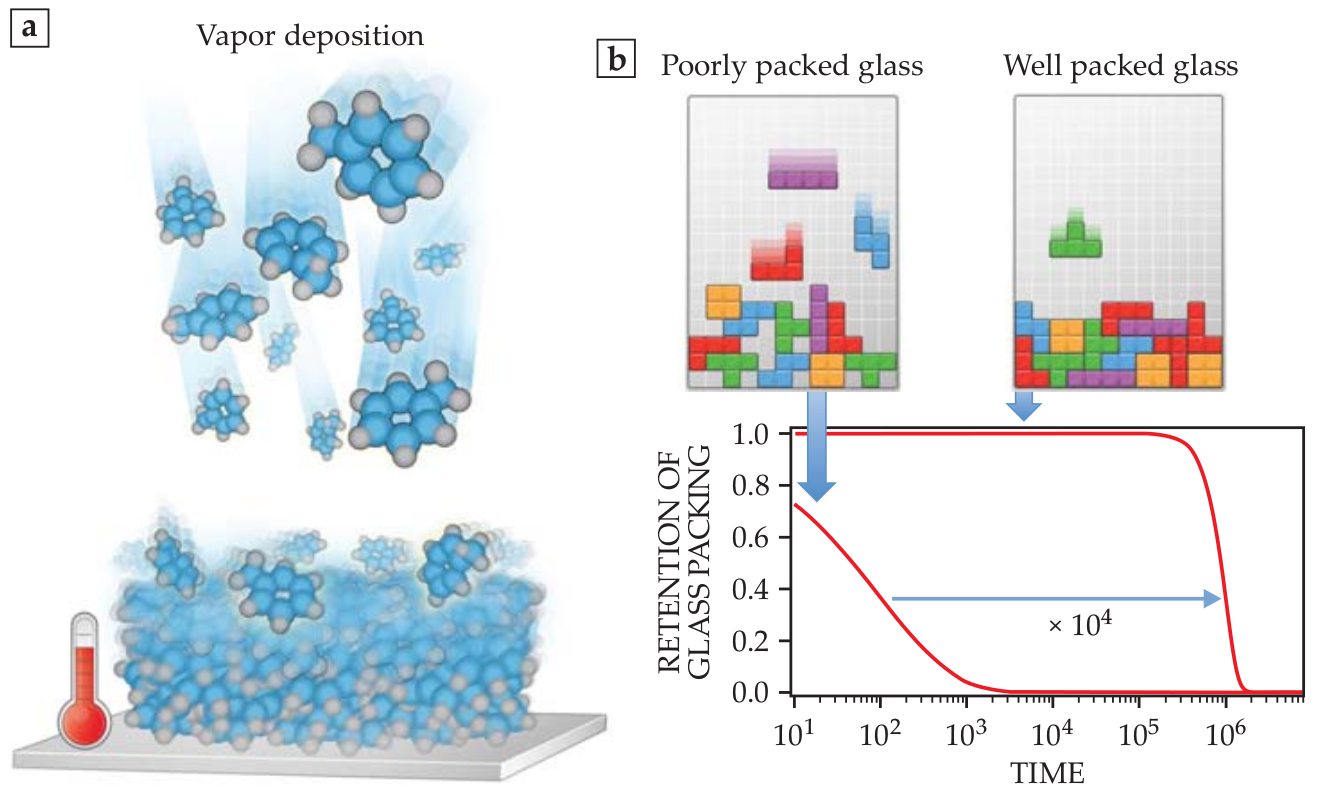}
\end{center}
\caption{{\bf Constructing better glasses using free surfaces.} (a) Glasses
produced by vapor deposition can exhibit highly efficient packing. 
Molecular mobility near the free surface is key to producing a well-packed
glass. The incoming molecules rapidly equilibrate if the substrate 
temperature is chosen correctly. (b) Block-stacking video games give an 
intuitive feel for how a slower deposition rate allows for better glass
packing. When a glass is heated to temperatures greater than the glass
transition temperature, molecular rearrangements start to undo the
glass packing. The graph shows that a well packed vapor-deposited
glass has dramatically increased kinetic stability; retention of glass
packing can be determined from density, enthalpy, or optical 
properties. (Illustrations courtesy of Hannah Sandvold.) }
\label{fig3}
\end{figure}

Physical vapor deposition produces high-density glasses with
remarkable properties. They have lower enthalpy than liquid-cooled 
glasses and provide the first indications of how supercooled
liquids might behave if the liquid state could be extended to 
temperatures below $T_g$. The materials have higher stiffness and their
packing can be so efficient that they transform into a liquid via a
sharp transformation front when heated above $T_g$. That behavior is
more like the isothermal melting of a crystal than the gradual 
softening observed in liquid-cooled glasses.

In another indication of high-density glasses’ efficient packing,
their heat capacities were recently shown to maintain a cubic temperature 
dependence~\cite{tls} down to $0.6$~K. In that regard, high-density
glasses behave like nonmetallic crystals in which phonons are the
dominant contributors to heat capacity. In contrast, the heat capacities 
of liquid-cooled glasses show a roughly linear temperature dependence 
at low temperatures, a behavior that had been interpreted
as evidence for universal low-temperature excitations in amorphous solids.
The deposition conditions that produce high-density glasses
also can produce oriented glasses in which the molecules adopt, for
example, planar orientation in the film. Organic LEDs, used in
many millions of mobile phone displays, are glasses produced by
vapor deposition. Creating planar orientation of the emitting 
molecules in those glassy films could increase the display efficiency by
more than 30~\%.

\section*{A genuine state of matter?}

In practice, glasses prepared from liquids use finite cooling rates 
and form by falling out of equilibrium with respect to the supercooled liquid.
What state would hypothetically result if a liquid could be
cooled infinitely slowly without crystallization? Can an equilibrium
liquid-to-glass phase transition exist? Those questions touch on
fundamental issues in the statistical mechanics of phase transitions
for complex systems containing disorder, impurities, and many-body 
interactions. Despite decades of intense research and steady
progress, the questions have not yet been answered 
satisfactorily~\cite{mark_review,rmp}.

In the conventional Landau approach to phase transitions, one
must first identify an order parameter (for example, the magnetization in 
ferromagnetic transitions or the density in liquid-gas transitions), before 
surmising an expression for the free energy based
on general symmetry considerations. For first-order phase transitions, 
the order parameter discontinuously becomes nonzero. For
second-order phase transitions, it obeys algebraic scaling laws and
goes continuously to zero near critical points.
For the putative liquid-to-glass transition, the choice of an order
parameter is not at all obvious because the molecular arrangements
in the glass are so similar to the ones found in the liquid (see Figure
2). There is no obvious symmetry breaking between the two states.
Rather, they are dynamically distinguished by the fact that
liquids can freely explore many different disordered configurations
while glasses cannot.

The glass-transition problem was recently solved in the mean-field limit, 
in a way that corresponds roughly to the Curie-Weiss
approach to ferromagnetism, or the van der Waals theory of the liquid-gas 
transition, both of which neglect fluctuations. Specifically,
the exact solution for the equilibrium phase diagram of hard 
spherical particles interacting in $d$ spatial dimensions was rigorously 
obtained in the mathematical limit $d \to \infty$, where mean-field 
approximations become exact~\cite{fractal}. 
In that limit, one can firmly establish the
existence of an equilibrium phase transition between liquid and
glass states and study its nature in full detail. The resulting van der
Waals picture of the glass transition is deeply rooted in theoretical
developments on phase transitions in disordered materials going
back at least 25 years~\cite{rfot}.

\begin{figure}
\begin{center}
\includegraphics[width=13cm]{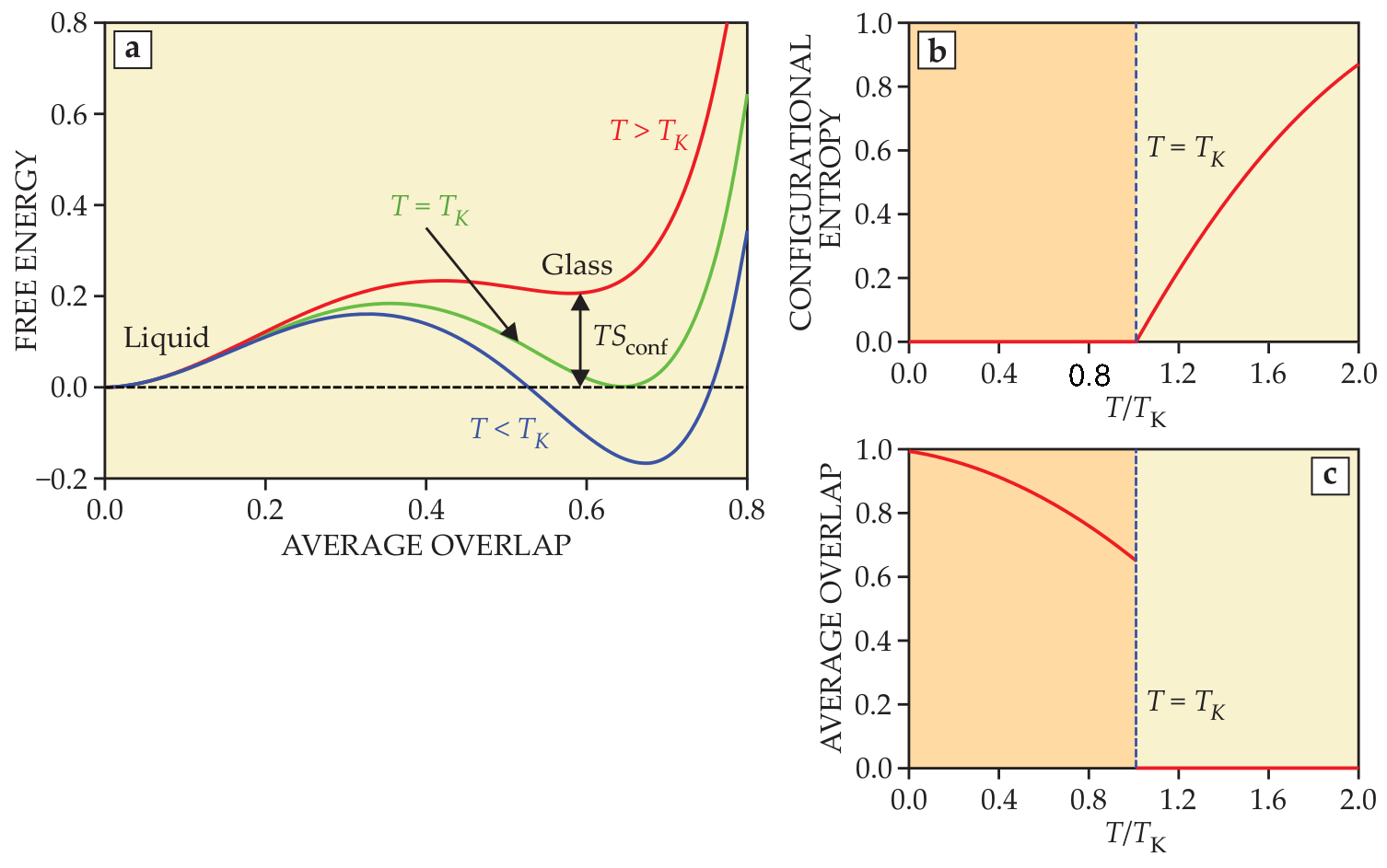}
\end{center}
\caption{{\bf Mean-field theory of the equilibrium liquid-to-glass phase
transition.} (a) The equilibrium free energy is shown in a representation
analogous to the liquid-gas transition, but with the average overlap,
which quantifies the similarities in pairs of configurations, in place of
density on the horizontal axis. The liquid (the state with zero average
overlap) is the stable phase at high temperature. (b) As temperature
decreases toward the so-called Kauzmann transition temperature $T_K$,
the configurational entropy $S_{\rm conf}$ (related to the free-energy difference
between liquid and glass states) decreases rapidly and vanishes at $T_K$.
(c) At that point, the system jumps discontinuously to the glass phase
with high average overlap. Below $T_K$, the glass state is the 
thermodynamically stable phase, the overlap is large, and the 
configurational entropy is zero.}
\label{fig4}
\end{figure}

In the mean-field theory of the simpler liquid-gas problem, two
minima in the free energy correspond to liquid and gas states. The
two states are readily distinguished by their density, which serves
as the order parameter. However, glass and liquid states are 
structurally too close for the glass transition to be treated in the same
way. In place of density, glass physicists have to construct a novel
object, called the overlap function $Q$, to distinguish the glass and
liquid states. The function describes the degree of similarity 
between the molecular positions in statistically independent pairs of
configurations equilibrated at the same temperature; $Q$ ranges from
$0$ for no similarity to $1$ for identical. One can then calculate in the
mean-field limit the free energy $V(Q)$ expressed in terms of the 
average overlap~\cite{rfot}. As shown in Figure 4, the free energy again has two
minima. The minimum at $Q = 0$ corresponds to the liquid, and the
minimum at high $Q$ corresponds to the glass.
If a pair of configurations is randomly drawn from the liquid state
at high temperature, so many configurations are available that 
particle positions are almost sure to be very different from one another.
Thus the $Q = 0$ minimum dominates the free-energy landscape. In
contrast, at low temperatures, two independent glass 
configurations are nearly identical (but still disordered), and their mutual
overlap $Q$ is large. The temperature at which the two free-energy
minima possess the same value is called the Kauzmann 
temperature $T_K$. Within mean-field theory, the ideal glass transition appears
as a discontinuous change in $Q$ at $T_K$ that separates a low-overlap
phase at high temperature and a high-overlap phase at low 
temperature, and for that reason it was dubbed a random first-order 
transition~\cite{rfot}. 

Above $T_K$, the system must pay a free-energy cost if it is to 
occupy a restricted part of its potential-energy landscape. The 
free-energy difference between the stable liquid and metastable glass
states quantifies that cost. Preventing the system from exploring
different states entails an entropic loss, called the configurational
entropy $S_{\rm conf}$. In mean-field theory, $S_{\rm conf}$ 
vanishes as temperature 
decreases toward $T_K$, as shown in Figure 4, and it remains zero in the
glass phase.

The concept of an ideal glass transition associated with a 
vanishing $S_{\rm conf}$ goes back almost 70 years to work by Kauzmann
and by Adam and Gibbs~\cite{kauzmann}. The recent production of
vapor-deposited ultrastable glasses described above suggests that
there might be ways to directly probe supercooled liquids with low
$S_{\rm conf}$ in real materials. Additionally, researchers are developing
novel computational techniques to explore experimentally relevant
temperature regimes and provide more direct insight into the 
validity of mean-field theory in finite dimensions~\cite{pts}.

For the liquid state above $T_g$, the connection between the 
thermodynamic picture and slow molecular motion is more 
complicated to establish and is thus more controversial. However, the 
analogy with first-order transitions suggests a physical mechanism to
explain the dynamics of molecular rearrangements.
In Figure 4, $S_{\rm conf}$ above $T_K$ represents the thermodynamic driving
force for relaxation from the localized, metastable glass state to the
liquid phase. Scaling arguments inspired by classical nucleation
theory provide a time scale for relaxation that grows exponentially
with $1/(k_B T S_{\rm conf})$, where $k_B$ denotes the Boltzmann constant. That
time scale diverges as $T_K$ is approached from above and $S_{\rm conf}$ goes
to zero, and it is infinite in the entire glass phase below $T_K$. The
physical interpretation is that dynamics slow as temperature 
decreases because ever fewer configurations are available, and the 
kinetic pathways between them become more complex and more collective.

Past research on the physics of phase transitions has repeatedly
warned us that ideas that appear pertinent in the mean-field limit
may break down completely when finite dimensional fluctuations
are included. Whereas such fluctuations have been successfully 
included in modern theoretical descriptions of simple phase transitions, 
they are still being intensely studied for problems where 
disorder and complex free-energy landscapes appear.
In parallel, researchers are exploring theoretical perspectives 
alternative to the mean-field approach, for instance those based on
real-space dynamic excitations that describe the liquid relaxation
dynamics via the emergence of a sparse collection of spatially 
correlated molecular displacements~\cite{chandler}. Another approach considers 
locally favored geometrical motifs and analyzes the overall 
disordered liquid structure as an assembly of topologically distinct
clusters~\cite{gilles}.

\section*{Do glasses contain defects?}

Defects are central to the physics of ordered condensed-matter
phases. Physicists typically view deformation and fracture in 
crystalline solids in terms of defect dynamics. For example, crystalline
metals can deform when one plane of atoms slides over another.
Given the intrinsic disorder of glassy materials, analogous slip
planes can obviously not occur in glasses.
The whole idea of defects in glasses might be dismissed by 
saying that everything about the packing in glasses appears defective,
whereas defects are useful objects only when they are sparse. 
Surprisingly, recent research has identified sparse defects or soft spots
that seem to play an important role.

When a glass is deformed at low temperature, one observes 
spatially localized irreversible rearrangements, often called shear
transformation zones, even at low strains~\cite{argon}. Shear transformation
zones act as defects but are only revealed by mechanical 
deformations. Indeed, they do not relate to obvious structural features, 
unlike defects in crystals, and that makes it hard to predict where
glasses will begin to flow. In computer simulations or in 
experiments that track colloidal particles with a microscope, shear 
transformation zones are easy to find, but only after the fact.
The current view is that when a glass is deformed, the strain
first becomes localized in a small number of shear transformation
zones. On further deformation, those zones organize spatially to
initiate shear bands that represent planes where the entire 
deformation of the material concentrates. That concentration eventually
leads to the large-scale failure of the glass.

Because shear transformation zones provide a molecular view
of how glasses break, understanding their structural origin and 
predicting their behavior are important research goals. Based on the
results of recent computer simulations, researchers have 
hypothesized that shear transformation zones 
in glasses are correlated with localized low 
frequency vibrations~\cite{defects}.
In a crystal, the lowest-frequency vibrational modes involve a
large number of atoms. Although some low-frequency modes in
glasses also have that character, additional low-frequency modes
involve an unusually small number of atoms. Such modes do not
exist in perfect crystals. It is those localized low-frequency modes
that appear to correlate with shear transformation zones. One might
think of shear transformation zones as being unusually poorly
packed regions of the glass, but at present no structural measure 
allows for their direct identification.

As the temperature is raised, thermal fluctuations 
spontaneously trigger more and more molecular rearrangements. 
Experiments and simulations indicate that just above $T_g$, molecular motion
is spatially heterogeneous and temporally highly intermittent. In
other words, some regions of the glass are mobile, whereas others
are immobile, and a given region might undergo long periods of
immobility separated by bursts of activity. Such dynamic 
heterogeneity~\cite{physics} 
means that thermally activated flow is highly collective
near $T_g$ and is characterized by a dynamic correlation length that
grows modestly as the temperature is lowered. It is tempting to 
envision a connection between the more mobile regions and the shear
transformation zones observed during flow in the glass. Elucidating
such a connection is also a topic of current research.

\section*{Glass physics in other areas of science}

Glassiness and the physics of arrested states with disordered 
structures are relevant for a wide range of materials with different length
and time scales. Figure 5 highlights several examples from soft 
condensed-matter physics and biology. 

\begin{figure}
\begin{center}
\includegraphics[width=15cm]{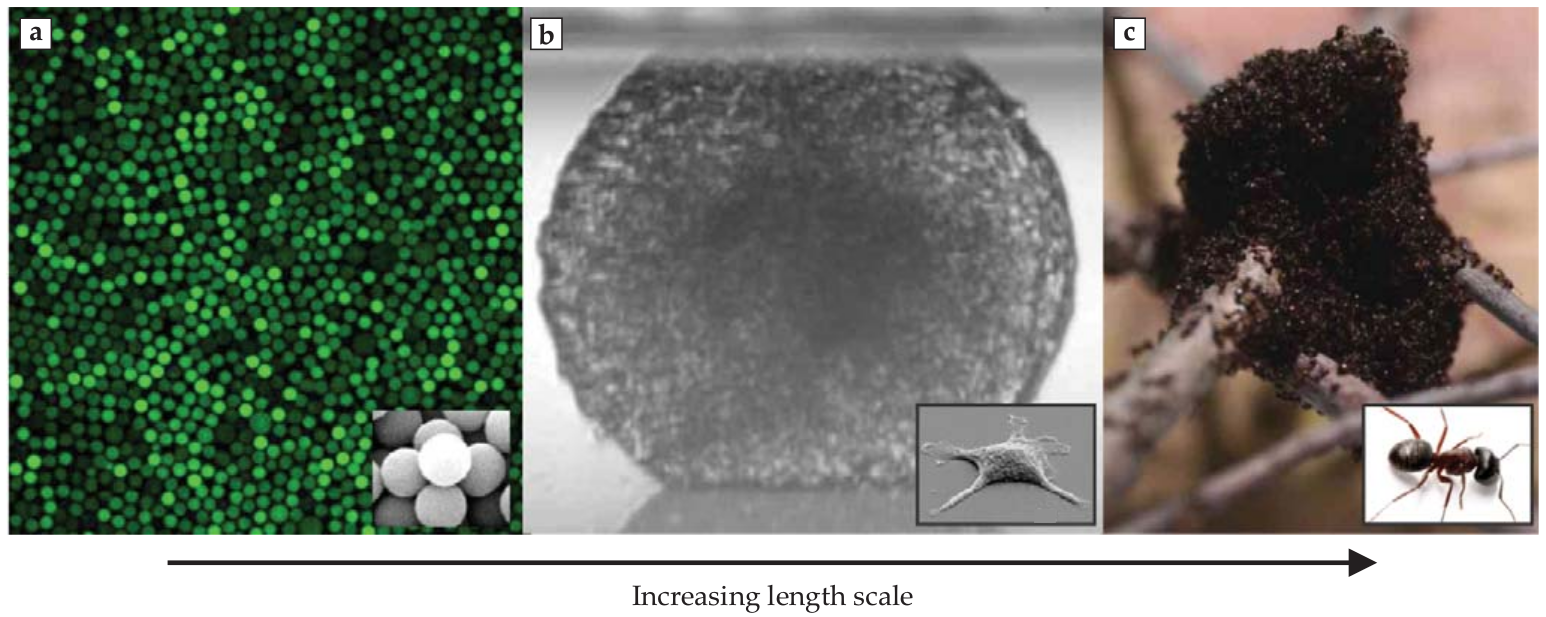}
\end{center}
\caption{Glassy physics at many length scales. Atomic liquids and
systems composed of larger particles are analogous in many ways. (a)
Highly concentrated colloidal micrometer-sized spheres resemble the
liquid and glass states shown in Figure 2. (b) A dense viscous droplet
composed of living cells squeezed between two plates exhibits many
glassy properties. (c) A dense swarm of ants retains its shape for a long
time, in analogy with an amorphous solid. (Panel a courtesy of Arran
Curran; panel b adapted from T. V. Stirbat et al., Eur. Phys. J. E 36, 84,
2013; panel c adapted from M. Tennenbaum et al., Nat. Mater., in press,
doi:10.1038/nmat4450.)}
\label{fig5}
\end{figure}

Soft materials are often dense
assemblies of supramolecular objects. Glasses made of colloidal
particles that range in size from 20 nm to 1 $\mu$m serve as simplified
model systems for studying the glass transition. The colloidal 
particles act as giant atoms that are large enough to be seen in a 
microscope but small enough to exhibit substantial Brownian motion
when suspended in a solvent. Experiments on colloidal glasses are
important links between computer simulations of simplified 
models and experimental investigations of molecular glasses~\cite{weeks}.
Because colloidal interactions can be tuned by various physical
means, colloidal glasses can be formed with a great variety of 
microscopic interactions. Colloidal particles with very short-range 
interactions, for instance, behave as sticky spheres and offer a model 
system that is unique to the colloidal world. The
vibrational and mechanical properties of these novel soft glassy 
materials differ qualitatively from glasses made with atoms or 
molecules that have longer-ranged interactions relative to the particle
size~\cite{weeks}.

Colloidal particles with sticky patches might allow detailed 
control of particle bonding along preferred angles, the first step to 
synthesize colloidal analogs of a broad variety of glasses, among them
silicon dioxide, the main component of window glasses, and glassy
water, whose properties remain debated. Such novel colloidal 
materials could then be used to visualize and understand, at the 
particle scale, the physical properties of their molecular counterparts.

Researchers are also increasingly studying glass transitions in
the framework of active materials, such as self-propelled colloidal
particles. Those studies have promising connections to biological
systems, including dense bacterial colonies and collective cell 
dynamics in epithelial tissues~\cite{active}. In many biological systems, the 
density of particles can be large, and consequently, the microscopic 
dynamics can become slow and glassy.
To tackle questions related to the mechanical properties of
dense tissue or to the diffusion of proteins in the crowded 
environment of cells, our understanding of glass physics is being expanded
to include chemical and mechanical driving forces acting in 
addition to thermal fluctuations. Human crowds and animal colonies
are also examples of materials that easily undergo dynamic arrest:
When the density becomes too large, flow almost ceases: think
about how difficult it is to move in the subway at rush hour.

Glass transitions also occur in more abstract problems in 
computer science~\cite{marc}. One example is an optimization algorithm that
seeks to answer a set of questions given a set of constraints. As the
number of constraints becomes large, satisfying them all becomes
increasingly difficult. The computer can get lost in a large number
of good-but-imperfect answers. Consider, for instance, the packing
problem: The task is to fill a box with hard, nonoverlapping objects.
When the density of objects is large, the constraint that no two 
objects can overlap becomes so demanding that one stops finding 
possible solutions above a suboptimal density. That situation is 
reminiscent of the glass transition.

A problem is deemed computationally hard if it cannot be
solved in a number of steps that is a polynomial function of the
number of inputs. For some classes of hard computational 
problems, methods invented to treat the statistical mechanics of glasses
have shown that the number of solutions decreases when the 
number of constraints increases in a way that is directly related to the
liquid-to-glass transition shown in Figure 4. Such methods have
even led to the development of physics-inspired computational 
algorithms for solving hard problems in computer science. The 
algorithms are successful precisely because they are based on methods
devised to tackle the complex free-energy landscapes that glass 
scientists face~\cite{marc}. Those methods now find applications in domains
such as image-compression and error-correcting codes.

Glass physics is rich and broadly applicable, and the above 
examples show that essential ideas from glass science are influencing
fields ranging from soft condensed-matter physics and biophysics
to computer science. Along with finding new uses for glassy 
materials as essential components in many modern technologies, 
researchers continue to make significant experimental and theoretical
progress toward a satisfying fundamental understanding of the
glass state.

\ack
Ludovic Berthier acknowledges support from the European Research
Council. Mark Ediger acknowledges support from NSF and the 
Department of Energy, Basic Energy Sciences, and thanks Reid Cooper for 
helpful discussions.

\end{document}